\documentstyle[epsfig,prl,aps,tighten,twocolumn]{revtex}

\newcommand{\be}{\begin{equation}}
\newcommand{\ee}{\end{equation}}
\newcommand{\bea}{\begin{eqnarray}}
\newcommand{\eea}{\end{eqnarray}}
\newcommand{\ba}{\begin{array}}
\newcommand{\ea}{\end{array}}

\begin{document}
\draft
\title{{\bf Spatiotemporally localized solitons in resonantly absorbing 
Bragg reflectors}}
\author{M. Blaauboer,$^{{\rm a,b}}$ G. Kurizki,$^{{\rm a}}$ 
and B.A. Malomed$^{{\rm b
}}$}
\address{
$^a$ Chemical Physics Department, Weizmann Institute of Science,\\
Rehovot 76100, Israel\\
$^b$ Department of Interdisciplinary Studies, Faculty of Engineering,\\
Tel Aviv University, Tel Aviv 69978, Israel}
\date{\today}
\maketitle

\begin{abstract}
We predict the existence of spatiotemporal solitons (``light bullets'') in 
two-dimensional self-induced transparency media embedded in a 
Bragg grating.
The ``bullets" are found in an approximate analytical form, their 
stability being confirmed by direct
simulations. These findings suggest new possibilities for signal
transmission control and self-trapping of light.
\end{abstract}


Light propagation in periodic dielectric structures exhibits 
a variety of unique regimes that 
are technologically promising: nonlinear filtering, 
switching and distributed feedback amplification\cite{dowl00}. Of particular 
interest are gap solitons (GSs), i.e., 
moving or standing self-localized field structures
centered in a band gap of the grating. These self-localized field structures
arise due to the interplay between the medium nonlinearity and resonant Bragg
reflections. Their spectrum is tuned away from the Bragg resonance by the 
nonlinearity at sufficiently high field intensities. Theoretical studies
of gap solitons in Bragg gratings with 
Kerr nonlinearity\cite{acev92} have been followed 
up by their experimental observation in a nonlinear optical fiber with the grating
written on it\cite{eggl96}. Gap solitons have also been theoretically studied in
gratings with second harmonic generation\cite{pesc97}. 

\begin{figure}[tbp]
\centerline{\epsfig{figure=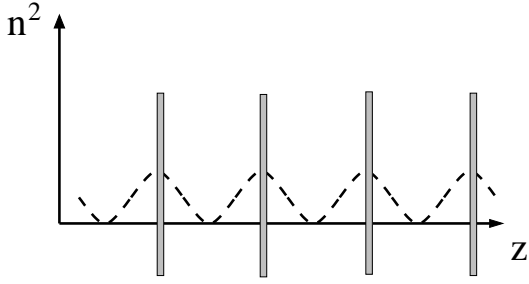,width=0.8\hsize}\vspace{0.5cm}}
\caption[]{Periodic modulation~(\ref{eq:perrefr}) of the refractive index 
in a structure of periodically alternating layers. Adapted from Ref.~\cite{kozh95}.
}
\label{fig:RABRschem}
\end{figure}

A principally different mechanism giving rise to gap solitons has been found in
studies of models consisting of a 
periodic array of thin layers of {\it resonant two-level systems} (TLS's)
separated by half-wavelength nonabsorbing dielectric layers, i.e., a
{\it resonantly absorbing Bragg reflector} (RABR)\cite{kozh95,kozh98,opat99}
(Fig. \ref{fig:RABRschem}). Such a RABR
has been shown, for {\it any} Bragg reflectivity, to produce a vast family
of stable solitons, both standing and moving ones. 
As opposed to the 2$\pi$ solitons
arising in self-induced transparency (SIT), 
i.e., near-resonant field-TLS interaction 
in a {\em uniform medium\/}\cite{mcca67,newe92,maim90}, 
gap solitons in a RABR can have an 
{\it arbitrary} pulse area\cite{kozh95,kozh98,opat99}. 
The existence of GS solutions can only be 
consistently demonstrated in a RABR with {\it thin} active TLS layers.
By contrast, a recent attempt\cite{akoz98} to obtain such solutions in a 
periodic structure {\it uniformly} filled with active TLS's is physically
doubtful, and fails for many parameter values. 

The potential applications
of GSs are based on the system's ability to filter out (by Bragg 
reflection) all pulses except for those satisfying the GS dispersion
condition, as well as the control 
of pulse shape and velocity\cite{kozh95,kozh98,opat99}. 
It would be
clearly desirable to supplement these advantages by immunity to transverse
diffraction of the pulse, i.e. to achieve {\it simultaneous} transverse 
as well as longitudinal self-localization in the structure. 
This calls for the consideration 
of ``light bullets'' (LBs), 
multi-dimensional solitons that are localized in both
space and time. In the last decade they have been theoretically investigated
in various nonlinear optical media
\cite{silb90,haya93,He98,kher98,fran98,gott98,drum84}, 
and the first experimental
observation of a quasi two-dimensional (2D) bullet was recently reported\cite
{liu99}. In a recent work~\cite{blaa00}, we have predicted that uniformly 
2D and 3D self-induced transparency (SIT) media can support stable light bullets.

In the present work, we aim to extend this investigation to RABR's
of the kind shown in Fig.~\ref{fig:RABRschem}, 
so as 
to combine  LB and GS properties in resonantly absorbing media. We
find that a RABR with {\it any} Bragg reflectivity 
and  {\it any} absorption cross-section can
support the propagation of attenuated
stable LBs, which are closely related to the bullets 
we have found in uniform SIT media 
\cite{blaa00}. It should be noted that 2D\ LBs supported by a
combination of the Bragg reflector with a \ different
(second-harmonic-generation) nonlinearity were theoretically investigated in
Ref. \cite{He98}.

We start by considering a 2D SIT medium with a spatially-varying refractive
index $n(z,x)$, which is described by the lossless Maxwell-Bloch equations 
\cite{newe92}, 
\begin{mathletters}
\label{eq:SIT1}
\begin{eqnarray}
-i{\cal E}_{xx}+n^{2}\,{\cal E}_{\tau }+{\cal E}_{z}+i\,(1-n^{2})\,{\cal E}-%
{\cal P} &=&0,  \label{eq:SIT11} \\
{\cal P}_{\tau }-{\cal E}W &=&0,  \label{eq:SIT12} \\
W_{\tau }+\frac{1}{2}({\cal E}^{\ast }{\cal P}+{\cal P}^{\ast }{\cal E})
&=&0.  \label{eq:SIT13}
\end{eqnarray}
\end{mathletters}
Here ${\cal E}$ and ${\cal P}$ are the slowly varying amplitudes of the
electric field and polarization, $W$ is the inversion, $z$ and $x$ are the
longitudinal and transverse coordinates (measured in units of the
resonant-absorption length), and $\tau $ is time (measured in units of the
input pulse duration). The Fresnel number, which governs the transverse
diffraction in the 2D and 3D propagation, has been incorporated in $x$ and
the detuning of the carrier frequency $\omega _{0}$ from the central
atomic-resonance frequency was absorbed in ${\cal E}$ and ${\cal P}$.
The Fresnel number $F$, detuning $\Delta \Omega $, and
wave vector $k_{0}\equiv \omega _{0}/c$ can be brought back into Eqs.~(\ref
{eq:SIT1}) by the transformation ${\cal E}(\tau ,z,x)\rightarrow \left(
2/k_{0}\right) \,{\cal E}(\tau ,z,x)\,\mbox{\rm exp}(-i\Delta \Omega \cdot
\tau )$, ${\cal P}(\tau ,z,x)\rightarrow \left( 4/k_{0}^{2}\right) \,{\cal P}%
(\tau ,z,x)\,\mbox{\rm exp}(-i\Delta \Omega \tau )$, $W(\tau
,z,x)\rightarrow \left( 4/k_{0}^{2}\right) \,W$, $\tau \rightarrow \left(
2/k_{0}\right) \,\tau $, $z\rightarrow \left( 2/k_{0}\right) \,z$, and $%
x\rightarrow \sqrt{2/k_{0}F}\,x$. To neglect the polarization dephasing and 
inversion decay we assume
pulse durations that are short on the time scale of the relaxation
processes. Equations~(\ref{eq:SIT1}) are then compatible with the local
constraint $|{\cal P}|^{2}+W^{2}=1$, which corresponds to the Bloch-vector
conservation \cite{newe92}. In the absence of the $x$-dependence and for $
n(z,x)=1$, Eq.~(\ref{eq:SIT11}) reduces to the sine-Gordon (SG) equation
which has the soliton solution ${\cal E}(\tau ,z)=2\alpha \,\mbox{\rm sech}
(\alpha \tau -z/\alpha +\Theta _{0})$, where $\alpha $ and $\Theta _{0}$ are
real parameters.
We proceed to search for LBs in a 2D medium subject to a resonant
periodic longitudinal modulation, i.e., the above-mentioned
RABR (Fig.~\ref{fig:RABRschem}). Accordingly, we assume the following periodic 
modulation of the refractive index\cite{series} 
\begin{equation}
n^{2}(z)=n_{0}^{2}\,\left[ 1+a_{1}\cos (2k_{c}z)\right] .
\label{eq:perrefr}
\end{equation}
Here, $n_{0}$ and $a_{1}$ are constants and $k_{c}=\omega _{c}/c$, with $%
\omega _{c}$ being the central frequency of the band gap. A RABR is then
constructed by placing very thin layers (much thinner than $1/k_{c}$) of
two-level atoms, whose resonance frequency is close to $\omega _{c}$, at the
maxima of this modulated refractive index. We aim to consider the propagation of an
electromagnetic wave with a frequency close to $\omega _{c} $ through a 2D
RABR. Due to the Bragg reflections, the electric field ${\cal E}$ is
decomposed into forward- and backward-propagating components ${\cal E}_{F}$
and ${\cal E}_{B}$, which satisfy equations that are a straightforward
generalization of the 1D version in Refs. \cite{kozh95,kozh98,opat99},
\begin{mathletters}
\label{eq:SIT4}
\begin{eqnarray}
-i\Sigma _{\tau xx}^{+}+i\Sigma _{zxx}^{-}+\Sigma _{\tau \tau }^{+}-\Sigma
_{zz}^{+} &&  \nonumber \\
+\eta \Sigma _{xx}^{+}+\eta ^{2}\Sigma ^{+}-2{\cal P}_{\tau }-2i\eta {\cal P}
&=&0,  \label{eq:SIT41} \\
-i\Sigma _{\tau xx}^{-}+i\Sigma _{zxx}^{+}+\Sigma _{\tau \tau }^{-}-\Sigma
_{zz}^{-} &&  \nonumber \\
-\eta \Sigma _{xx}^{-}+\eta ^{2}\Sigma ^{-}+2{\cal P}_{z} &=&0,
\label{eq:SIT42} \\
{\cal P}_{\tau }+i\Delta \Omega {\cal P}-\Sigma ^{+}W &=&0,  \label{eq:SIT43}
\\
W_{\tau }+\frac{1}{2}(\Sigma ^{+\ast }{\cal P}+\Sigma ^{+}{\cal P}^{\ast })
&=&0.  \label{eq:SIT44}
\end{eqnarray}
\end{mathletters}
Here $\Sigma ^{\pm }\equiv 2\tau _{0}\mu n_{0} ({\cal E}_{F}\pm {\cal E}%
_{B})/\hbar$, $\tau _{0}\equiv n_{0}\mu ^{-1} \sqrt{\hbar /2\pi \omega _{c}\rho
_{0}}$, $\mu $ is the transition dipole moment, and $\rho _{0}$ is the
density of the two-level atoms. The parameter $\eta $ is a ratio of
resonant-absorption length in the two-level medium to the Bragg reflection
length, and can be expressed as $\eta =a_{1}\omega _{c}\tau _{0}/4$. 

In the 1D case, a family of {\em exact} soliton solutions to Eqs.~(\ref{eq:SIT4})
was found in Ref.~\cite{kozh98}: $\Sigma^{(\pm)} = 2\left( \alpha ; -\sqrt{
\alpha^{2} + 2} \right) \mbox{\rm sech}\Theta (\tau ,z)\cdot \exp \left(
i\Phi \right) $, ${\cal P}=2\,\mbox{\rm sech}\Theta (\tau ,z)\,\tanh \Theta
(\tau ,z)\cdot \exp \left( i\Phi \right) $, $W=2\mbox{\rm sech}^{2}\Theta
(\tau ,z)-1$, where $\Theta (\tau ,z)\equiv \alpha \tau +\sqrt{\alpha ^{2}+2}%
\,z+\Theta _{0}$, $\Phi \equiv \eta M\tau +\eta Nz+\phi $, $M\equiv -(\alpha
^{2}+1)$, $N\equiv -\alpha \sqrt{\alpha ^{2}+2}$, and $\Delta \Omega =\eta
(\alpha ^{2}+1)$. The shape of the fields $\Sigma ^{+}$ and $\Sigma ^{-}$ in
this solution is similar to the SG soliton in the uniform 1D SIT medium.
Inspired by this analogy and the fact that there exist LBs in the uniform 2D
SIT medium which reduces to the SG soliton in 1D \cite{blaa00}, we search
for a LB solution to the 2D equations~(\ref{eq:SIT4}), which reduces to the
exact soliton in 1D. To this end, we try the following approximation, 
\begin{mathletters}
\label{eq:parm4}
\begin{eqnarray}
\Sigma ^{+} &=&2\,\alpha \,\sqrt{\mbox{\rm sech}\Theta _{1}\mbox{\rm
sech}\Theta _{2}}\ e^{i\eta M\tau +i\eta Nz+i\pi /4},  \label{eq:parm41} \\
\Sigma ^{-} &=&-2\,\sqrt{\alpha ^{2}+2}\,\sqrt{\mbox{\rm sech}\Theta _{1}%
\mbox{\rm sech}\Theta _{2}}\ e^{i\eta M\tau +i\eta Nz+i\pi /4},
\label{eq:parm42} \\
{\cal P} &=&\sqrt{\mbox{\rm sech}\Theta _{1}\mbox{\rm sech}\Theta _{2}}%
\{(\tanh \Theta _{1}+\tanh \Theta _{2})^{2}+  \nonumber \\
&&\frac{1}{4}\alpha ^{2}C^{4}[(\tanh \Theta _{1}-\tanh \Theta _{2})^{2}-2(%
\mbox{\rm sech}^{2}\Theta _{1}+  \nonumber \\
&&\mbox{\rm sech}^{2}\Theta _{2})]^{2}\}^{1/2}\ e^{i\eta M\tau +i\eta
Nz+i\nu },\,W=\left[ 1-|{\cal P}|^{2}\right] ^{1/2},  \label{eq:parm44}
\end{eqnarray}
with $\Theta _{1}(\tau ,z)\equiv \alpha \tau +\sqrt{\alpha ^{2}+2}\,z+\Theta
_{0}+Cx$, $\Theta _{2}(\tau ,z)\equiv \alpha \tau +\sqrt{\alpha ^{2}+2}%
\,z+\Theta _{0}-Cx$, the phase $\nu $ and coefficient $C$ being real
constants.
\end{mathletters}

\begin{figure}[tbp]
\centerline{\epsfig{figure=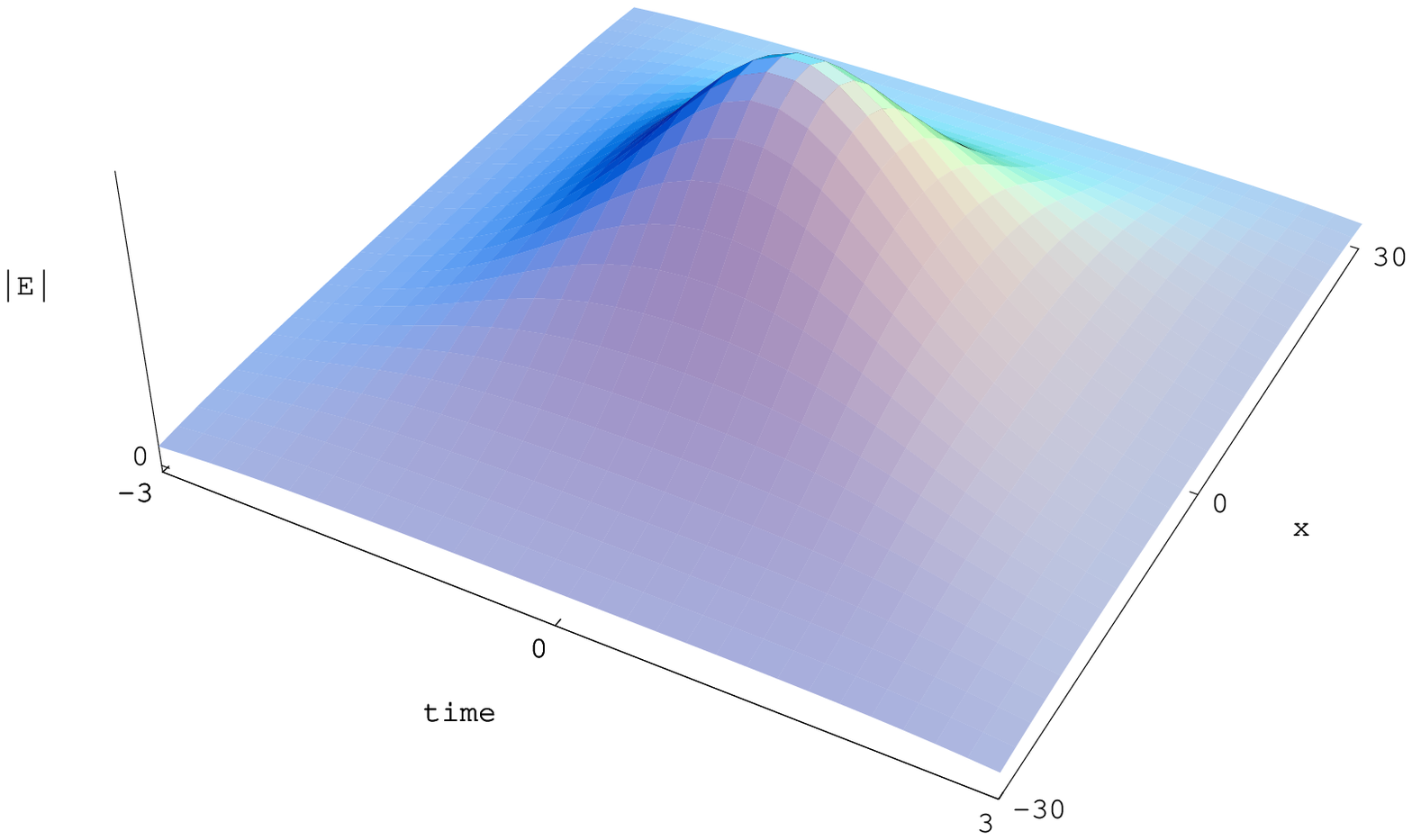,width=0.8\hsize}\vspace{0.5cm}}
\caption{The forward-propagating electric field of the 2D ``light bullet''
in the Bragg reflector, $|{\cal E}_{F}|$, vs. time $\protect\tau $ and
transverse coordinate $x$, after having propagated the distance $z=1000$.
The parameters are $\protect\alpha =1$, $C=0.1$ and $\Theta _{0}=-1000$. The
field is scaled by the constant $\hbar /4\protect\tau _{0}\protect\mu n_{0}$.
}
\label{fig:RABR}
\end{figure}

The ansatz~(\ref{eq:parm4}) satisfies~Eqs. (\ref{eq:SIT41}) and (\ref
{eq:SIT42}) exactly, while Eqs. (\ref{eq:SIT44}) are satisfied to order $%
|\alpha |C^{2}$, which requires that $|\alpha |C^{2}\ll 1$. The ansatz is
relevant for {\it arbitrary} $\eta $, admitting both weak ($\eta \ll 1$) and
strong ($\eta >1$) reflectivities of the Bragg grating, provided that the
detuning remains small with respect to the gap frequency, or $\eta \ll
\omega _{c}/(\alpha ^{2}+1)$. Comparison with numerical simulations of Eqs.~(%
\ref{eq:SIT4}), using (\ref{eq:parm4}) as an initial configuration, tests
this analytical approximation and shows that it is indeed fairly close to a
numerically exact solution; in particular, the shape of the bullet remains
within 98\% of its originally presumed shape after having propagated a large
distance, as is shown in Fig.~\ref{fig:RABR}. 

We stress that 2D or 3D LB solutions of the variable-separated form $\Sigma
^{+}\sim \Sigma ^{-}\sim f(\tau ,z)\cdot g(x)$ do {\em not} exist in RABR. Indeed,
substituting this into Eqs. (\ref{eq:SIT41}) and (\ref{eq:SIT42}) yields
only a trivial solution of the form $\Sigma ^{\pm }\sim \exp \left( iA\tau
+iB x\right) $, with constant $A$ and $B$.

We briefly discuss experimental conditions under which LBs
can be observed in RABRs. The incident pulse should have uniform
transverse intensity within its diameter $d$.
For the transverse diffraction to be strong
enough, one needs $\alpha _{{\rm eff}}d^{2}/\lambda _{0}<1$, where $\alpha _{%
{\rm eff}}$ and  $\lambda _{0}$  are the inverse resonant-absorption
length and  carrier wavelength, respectively\cite{slus74}. For $\alpha
_{{\rm eff}}\sim 10^{3}$\thinspace\ m$^{-1}$ and $\lambda _{0}\sim 10^{-4}$%
\thinspace\ m, one thus requires a diameter
$d<10^{-4}$ m, which implies that the
transverse medium size $L_{x}\sim \,$a few$\,\mu $m. 

In order to realize a RABR, thin layers of {\em rare-earth ions\/}
\cite{gre99} 
embedded in a semiconductor structure with a spatially-periodic RI 
\cite{khit99}
may be used. 
The two-level atoms in the layers should be rare-earth-ions with density 
of $10^{15}-10^{16}$\thinspace\ cm$^{-3}$, whose
resonant-absorption time and inverse length are
respectively $\tau _{0}\sim 10^{-13}-10^{-12}$\thinspace\ s and 
$\alpha _{{\rm eff}}\sim 10^{4}-10^{5}$%
\thinspace\ m$^{-1}$. The parameter $\eta $ can vary
from 0 to 100 and the detuning is $\sim 10^{12}-10^{13}$\thinspace\ s$^{-1}$%
. In a RABR with transverse size $10\,\mu $m, LBs depicted in Fig.~\ref
{fig:RABR} are localized on the time and transverse-length scales $\sim
10^{-13}$ s and $1\,\mu $m. Cryogenic conditions strongly extend the
dephasing time $T_2$ and thus the LB lifetime, well into the 
$\mu$sec range
\cite{gre99}. 
The construction of suitable structures constitutes a
feasible experimental challenge.

To conclude, we have studied light bullets in SIT\ media
embedded in a resonantly-absorbing Bragg reflector. 
Light bullets in a multi-dimensional Bragg reflector have the potential of
serving as a novel type of optical filters, which stably transmit selected
signal frequencies through their spectral gap and block others. They can
also be used to both spatially and temporarily localize light in certain
frequency bands.

M.B. acknowledges support through a fellowship from the Israeli Council for
Higher Education. Support from ISF, Minerva and BSF is acknowledged by
G.K.


\begin{references}
\bibitem{series}  In reality, $n^{2}(z)$ is represented by an infinite
Fourier series. For the propagation of waves whose wavevector is close to $%
k_{c}$, however, only the coefficient $a_{1}$ 
is essential (see Ref.~\cite{opat99}).
\end{references}
\end{document}